# Evidence for a quantum phase with macroscopic orbit-like quantum states similar to the current-carrying edging states in integer quantum Hall system

**Sergey A. Emelyanov**

*Ioffe Institute, 194021 St. Petersburg, Russia*

E-mail: sergey.emelyanov@mail.ioffe.ru

**Highlights**

● There is a quantum phase in which macroscopic quantum orbits fill the entire system

● The phase results from a quantum phase transition from integer quantum Hall system

● The transition is accompanied by the breaking of system translational symmetry

● A macroscopic system in the phase is a fully-correlated indivisible quantum object

● Such system is a unique object to test quantum foundations at a macroscopic level

**Abstract**

By the method of intense terahertz laser spectroscopy, we provide strong evidence that if an integer quantum Hall (IQH) system has asymmetric confining potential and the external quantizing magnetic field has a nonzero in-plane component, then a quantum phase may arise with spatially-ordered quasi-one-dimensional orbit-like states similar to the current-carrying edging states in conventional IQH system. The emergence of the phase may be interpreted in terms of a quantum phase transition from the IQH system, which is accompanied by the breaking of translational symmetry in the direction of the in-plane magnetic field. As a result, we get a fully-correlated macroscopic quantum object that gives one a unique opportunity to test the foundations of quantum mechanics (QM) at a macroscopic level. In particular, by means of this object, we show experimentally that the concept of quantum spatial dynamics of electrons without a definable trajectory remains relevant in the macro-world in accordance with the foundations of standard QM while an alternative description of such dynamics in terms of so-called Bohmian trajectories appears untenable.

**Graphical abstract:**

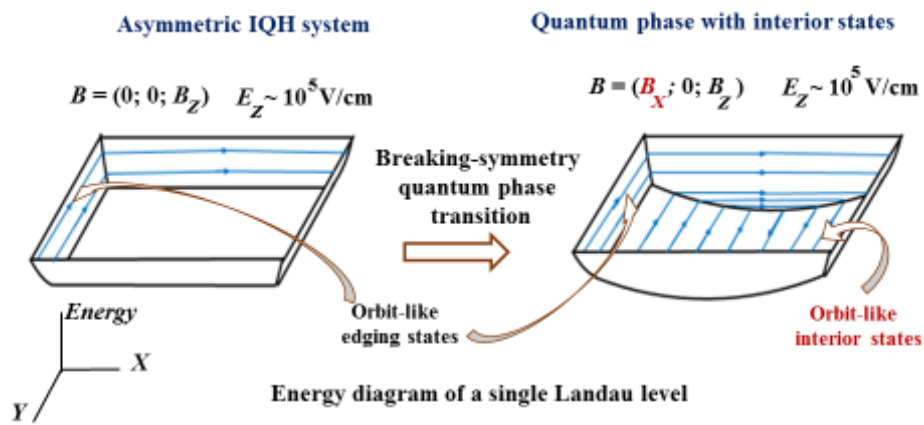

## 1. Introduction

Quantum Hall state is one of the low-energy states of matter, where purely quantum effects manifest themselves at a macroscopic level [1-3]. In many respects, this state of matter remains mysterious even up to now (see, e.g., [4]). One of the quantum Hall states is the so-called integer quantum Hall (IQH) state, the discovery of which by Klaus von Klitzing was awarded the Nobel Prize in 1985. Now the quantum properties of this state are applied in the field of metrology [5-6].

However, compared to the other low-energy states, the IQH state has an important feature that has a potentiality to be applied in a quite new field. The point is that here a macroscopic quantum behavior of electrons is not of a collective character but rather of an individual character. As a result, the IQH system could be a unique object to test the relevance of quantum foundations in the macro-world precisely on the level of individual quantum particles. But to date, such tests are hard to perform. This is due to the fact that the quantum states responsible for a macroscopic quantum behavior of the system are concentrated only in a microscopic proximity to the system edges and therefore most of the standard experimental methods appear inapplicable here to the exclusion of the well-known magneto-transport one

To clarify the problem, consider briefly what the IQH system is. As a rule, it is a semiconductor structure where the thickness of the conducting layer is less than the mean free path of free electrons. As a result, the electron motion perpendicular to the layer (along the $Z$-axis) is quantized and here their energy spectrum is a series of quantum levels. To achieve the IQH state, a strong magnetic field is additionally applied along the $Z$ axis. As a result, the electron motion in the $XY$ plane will also be quantized, where both $X$ and $Y$ axis are arbitrary because of the axial symmetry in the $XY$ plane. In fact, this means that each quantum level gives a series of sublevels, the so-called Landau levels, where electrons are strictly localized within their cyclotron orbits of a microscopic radius. Here the Landau levels are strongly degenerate with respect to the $X$-coordinate of the center of electrons' cyclotron orbits ($X_o$) or equivalently with respect to their wave vector in the $Y$ direction ($k_y$) since these two parameters are in a strict correlation: $X_o = -\lambda^2 k_y$, where $\lambda$ is the so-called magnetic length.

But the key point is that in a microscopic proximity to the edges of a real IQH structure, there is always a strong in-plane electric field perpendicular to the edges. As a result, here electrons are in a crossed electric and magnetic field and therefore, as was shown theoretically by Bertrand Halperin, macroscopic quasi-one-dimensional orbit-like edging states arise along the perimeter of the system [7]. Their characteristic cross-section is of the order of the cyclotron orbit and they are often called current-carrying edge states which are precisely the reason for the IQH effect [8-9]. Thus, if an IQH system is of the length $a$ ($X$-axis) and of width $b$ ($Y$-axis), then the energy diagram of each Landau level is something like a rectangular "energy bowl" with very thin walls where the orbit-like states are. Fig. 1 shows the energy diagram of such system where we neglect the local potential fluctuations.

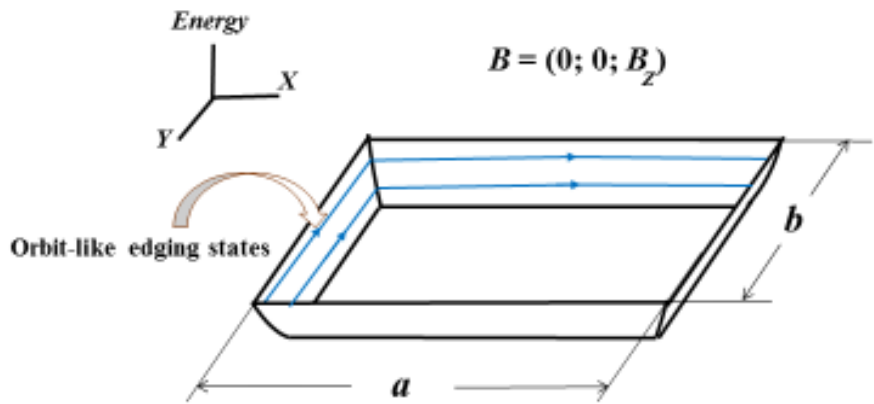

Fig. 1. The energy diagram of Landau level in conventional IQH system. The scale is strongly exaggerated near the edges to show the walls of the "energy bowl" with the macro-orbits related to a crossed electric and magnetic field.

Thus, the number of electrons responsible for macroscopic quantum behavior is much less than the number of localized electrons at the bowl bottom. Therefore, we have to apply an experimental method sensitive to the former but insensitive to the latter. Today only one such method is known. We mean the magneto-transport measurements used by von Klitzing because here the localized electrons can be neglected as they cannot contribute to a macroscopic transport. Today there are no other methods to deal with the macroscopic current-carrying edge states.

However, one would propose a way to avoid this difficulty through a modification of the IQH system itself. The idea rests on the fact that a crossed electric and magnetic field can be achieved not only near the edges but also in the system interior. The point is that the confining potential of a real IQH system may well be asymmetric for various reasons. This is an equivalent of the presence of an electric field along the Z-axis, which is often called the built-in field. Such field can be as high as up to $10^5$V/cm and, in particular, it is responsible for a strong spin-orbit coupling known as the Rashba effect [10-12]. Thus, if an asymmetric IQH system is in the external quantizing magnetic field with a nonzero in-plane component ($B_x$), then a crossed electric and magnetic field will be not only near the edges but also in the system interior.

Generally speaking, a situation of that kind was treated theoretically in the early 90s with a simplified model of endless system [13-14]. There it was shown that the transverse electric field in a combination with the in-plane component of magnetic field may truly lift the Landau level degeneracy throughout the system. As a result, Landau levels turn into a kind of energy bands, the minima of which are slightly shifted from each other in the X direction. In such a band, electrons behave as spontaneous currents along the Y-axis, which are spatially separated along the X-axis. The characteristic cross-section of such currents is of the order of the cyclotron orbit determined by $B_x$. But the most important thing is that each current to the left of the band minimum has a counterpart to the right of the minimum, which flows in the opposite direction. As a result, the total current along the Y-axis is always zero.

Now if we take a real asymmetric IQH system with a non-zero $B_x$, then it seems reasonable to assume that there will be a system of closed macro-orbits that fill the entire band. Fig. 2 shows the energy diagram of such band where we also neglect the local potential fluctuations. It is seen that the main difference of this diagram from the conventional one (see Fig. 1) is that the bottom of the energy bowl looks like a gutter extended along the Y-axis. Accordingly, the sides of the gutter are filled with the isoenergetic macro-orbits of Halperin type, which are extended along the Y-axis and closed near the edges parallel to the X-axis.

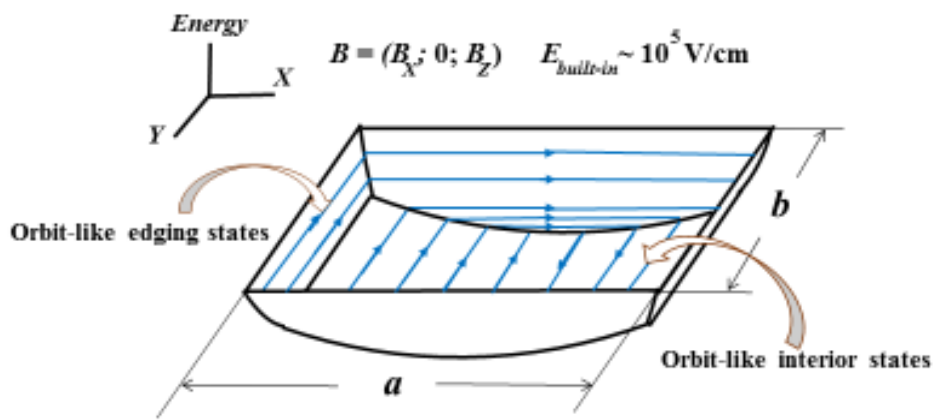

Fig. 2. The energy diagram of a Landau band in the IQH system with asymmetric confining potential in presence of a nonzero in-plane component of magnetic field. The degeneracy is lifted throughout the whole system and the Halperin-type macro-orbits fill the entire Landau band.

However, despite the fact that the calculations of Refs. [13-14] were performed many years ago, no evidence has been found so far for the spectrum in Fig. 2 by using of the standard experimental methods.

## 2. Experimental method

### 2.1 How to test the presence of quantum macro-orbits in the system interior?

An important point is that if the phase with quantum macro-orbits does exist, then it has a characteristic feature that could be detectable by means of an experimental method never applied to the IQH system. We mean the fact that the transition to this phase should be accompanied by the breaking of translational symmetry in the *X* direction. This is because the lifting of the Landau level degeneracy should be accompanied by a little shift of the newly-formed Landau bands along the *X*-axis and this shift should increase with increasing of the Landau quantum number (N). In this case, the energy diagram in the *X* direction for two adjacent Landau bands should look like that in Fig. 3. Here we assume that the electron density is high enough, and the temperature is much less than the energy gap between the bands, so that the N-th Landau band is fully occupied with electrons while the (N+1) band is vacant.

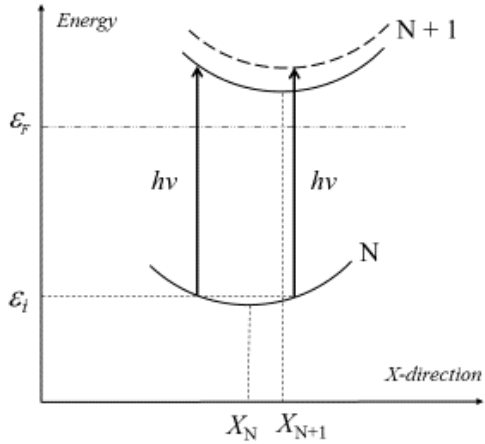

Fig. 3. An asymmetry of light-induced electron transitions between the neighboring Landau bands, which leads to an imbalance of the electronic system. The Fermi level ($\varepsilon_F$) is between the bands.

As seen from Fig. 3, if such system is excited by the laser radiation so that the energy of quanta is equal to the gap between the bands (the so-called cyclotron resonance or CR), then the *X*-coordinate of the CR point is a function of the quantizing magnetic fields. If we assume that the light-induced transitions occur between the macro-orbits, then the initial and the final states will not be completely identical. This may lead to an imbalance of electron fluxes along the *Y*-axis and, in presence of a quasi-elastic scattering, a non-zero local current may arise along the *Y*-axis, which is a function of both the external magnetic field and the *X*-coordinate.

Phenomenologically, it is the so-called photogalvanic effect, which, as a rule, occurs when terahertz laser radiation excites a semiconductor system with an inner asymmetry characterized by a polar vector. In this case, a photo-induced current may arise which is collinear to this vector and has a characteristic relaxation time of the order of momentum relaxation time of non-equilibrium electrons (for a review, see [15-16]). In our system, the asymmetry is related to non-zero vector product $B_x \times E_{built-in}$. But the key point is that if local photo-current in the *Y* direction is truly a function of the *X*-coordinate, then we are truly dealing with the breaking of translational symmetry in the *X* direction on a macroscopic lengthscale.

Actually, in some respects, the proposed experimental method resembles the von Klitzing's one. In particular, they both imply the measurement of macroscopic currents in the *XY* plane, which can only be related to the current-carrying macro-states. But the fundamental difference is that (1) our method is extremely sensitive to the system asymmetry associated with the shift of Landau bands (if any) and (2) now the measurable current is induced not by an electric bias but by a laser radiation. In this case, both the excitation of the system and the registration of its responses can be strictly local and this is especially important if we do deal with an indivisible quantum macro-system with strong quantum correlations between the its local regions

## 2.2 Source material and the experimental scheme

As a source material, we use a single quantum well structure of type GaSb-InAs-GaSb grown by the molecular beam epitaxy (MBE). The well width is 15 nm. The well of that type is known to be semi-metallic because the top of the GaSb valence band is above the bottom of the InAs conduction band. As a result, the electron density is relatively high even at low temperatures (about $1.5 \cdot 10^{12} cm^{-2}$). To avoid the so-called electron hybridization because of the band overlapping, the InAs quantum well is sandwiched between two thin layers of AlSb (3nm each). To achieve an asymmetry of confining potential, the GaSb cap layer is 20nm whereas the typical penetration depth of the so-called surface potential is much higher (about 100nm) [17]. The presence of such an asymmetry will be checked in the course of further experiments.

The structure is cooled down to 1.9K. In this case, the electron scattering is determined by charged point-like defects with a high density so that the scattering time is as short as about 3ps. As a result, the mean free path of free electrons is about $0.1\mu m$. Since the electron effective mass in InAs is relatively small, the energy of Landau quantization is close to the energy of terahertz laser quanta under the quantizing magnetic fields of less than 5T. For this reason, we use a superconducting magnet of up to 6T where magnetic field can be tilted in the *XZ* plane by an arbitrary angle depending on a concrete task.

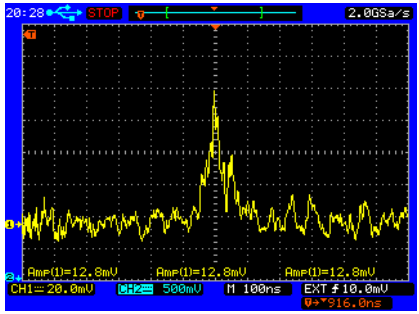

Fig. 4. Typical track of the terahertz laser pulse. Timescale is 100ns/div.

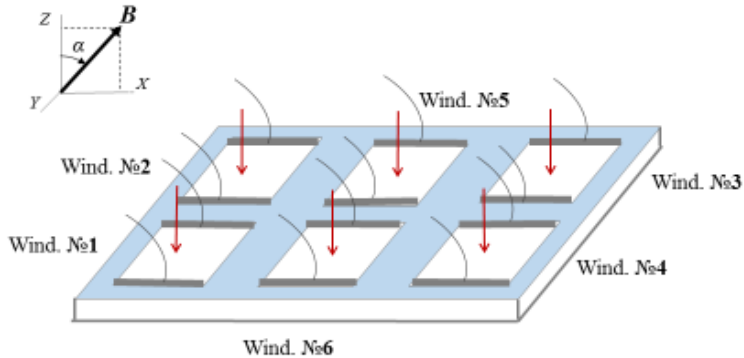

Fig. 5. Experimental scheme. The structure is covered by a non-transparent mask with six identical windows, through which terahertz laser radiation can enter the system. External magnetic field (*B*) can be tilted in the *XZ* plane by an arbitrary angle.

As a source of terahertz radiation, we use an ammonia laser optically pumped by $CO_2$ laser. The terahertz laser wavelength is $90.6\mu m$ ($h\nu = 13.7meV$). In this case, CR is expected at quantizing magnetic field of about 4.8 T. Since the in-plane system asymmetry is expected to be weak, the laser intensity is as high as about $200W/cm^2$. Therefore, to avoid heating effects, the duration of laser pulse is as short as about 50ns. Fig. 4 shows typical laser track recorded by the high-speed photon-drag detector [18]. The detection with a time resolution allows us to be sure that the kinetics of responses is similar to that of laser pulses and hence heating effects are neglectedly small.

Since by the measuring of local photo-currents we want to check whether or not the system translational symmetry is broken, we use the experimental scheme shown in Fig. 5. The structure is covered by a non-transparent plate with six identical windows through which laser radiation can enter the sample. The windows differ from each other only in their coordinates, so they can be regarded as a translation of a single window along both the *X*-axis and the *Y*-axis. The structure size is 21mm along the *X*-axis and 17mm along the *Y*-axis. Each window is 4 mm in length (*X*-axis) and 5 mm in width (*Y*-axis) and supplied with a pair of thin ohmic contacts (0.5 mm each) to measure local photo-currents in the *Y* direction. The distance between adjacent windows is 3 mm along the *X*-axis and 2 mm along the *Y*-axis. All windows are removed from the edges by no less than 1.5 mm to avoid edging effects.

3. **Results and discussion**

   3.1 **Evidence for translational symmetry breaking**

To avoid any ambiguity, all experiments will be done in two regimes that differ only in the direction of external magnetic field varied *in situ*. In the first one (regime *A*), magnetic field is parallel to the well plane so that there is no Landau quantization and we deal with free electrons of Bloch type in the *XY* plane. Since here the vector product $B_x \times E_{built-in}$ is also nonzero, a photo-current along the *Y*-axis may also arise and it was truly observed earlier [19]. But this current *a-priori* cannot be associated with macro-orbits.

In the second regime (regime *B*), we are near the CR point and the external magnetic field has a non-zero in-plane component. It is precisely the regime where we expect the breaking of translational symmetry in the *X* direction due to the internal macro-orbits. The results obtained in both these regimes will be compared with each other.

Fig. 6 shows the dependence of photo-current on the magnetic field in the regime *A* for two opposite directions of the field ($\alpha = \pm 90$). It is seen that, first, a photo-current does arise and it is proportional to the magnetic field. This fact confirms the presence of a built-in electric field in our structure. Secondly, photo-currents in all working regions are almost identical and they all are an odd function of magnetic field. This means that here translational symmetry remains. Finally, in presence of a high enough quantizing component ($\alpha = \pm 60$), photo-currents quickly disappear with increasing of magnetic field so that the mechanism of their arising cannot be relevant in the regime of Landau quantization.

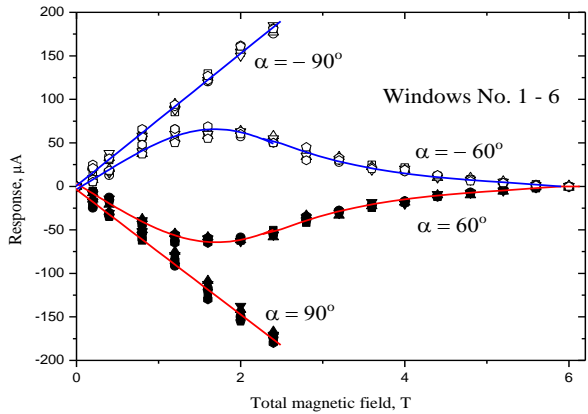

Fig. 6. The dependence of photo-currents on the total magnetic field in the regime *A*. Different symbols correspond to different working regions. Open symbols – negative angles (blue curves), closed symbols – positive angles (red curves). Solid curves are a guide for the eyes.

However, a completely different situation is observed in the regime *B*. Figs. 7a, 7b, and 8 show photo-responses in the working regions when all windows are open. It is seen that, first, here the same responses are observed only in the regions with the same *X*-coordinates. Otherwise, the responses may differ even in their sign. Moreover, this fact cannot be interpreted in terms of any local inhomogeneity because the behavior of responses is clearly sensitive to the direction of $B_x$. This means we are truly dealing with the breaking of translational symmetry in the *X* direction.

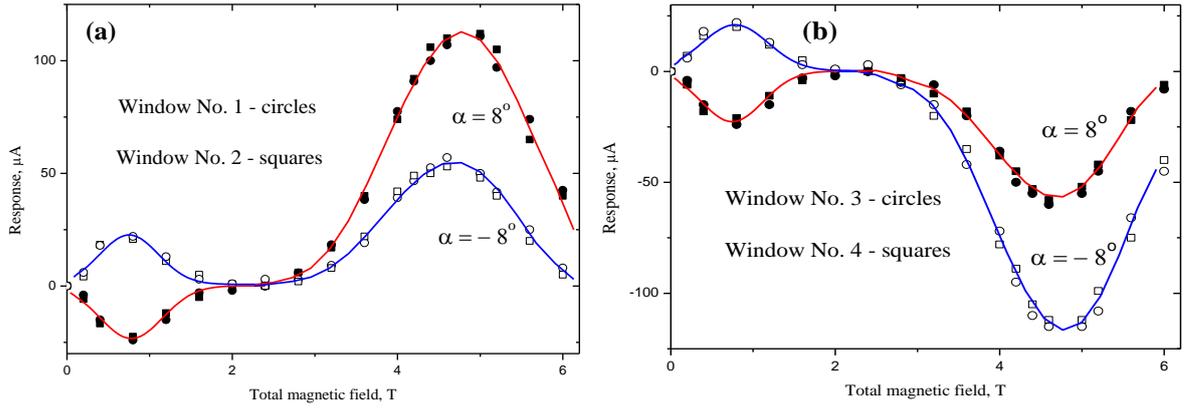

Fig. 7. The dependence of photo-responses on the total magnetic field in the regime *B*: (a) – windows No. 1 and No. 2, (b) – windows No. 3 and No. 4. Different symbols correspond to different working regions. Open symbols − negative angles (blue curves), closed symbols – positive angles (red curves).

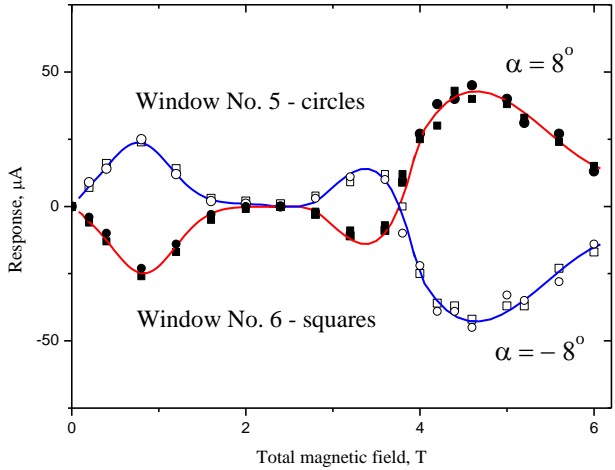

Fig. 8. The dependence of photo-responses on the total magnetic field in the regime *B* in the working regions No. 5 and No. 6.

Secondly, in some working regions, the responses are not an odd function of $B_x$. It is directly seen from their behavior in two pairs of windows: No. 1 and No. 2 (Fig. 7a) as well as No. 3 and No. 4 (Fig. 7b) where the switching of $B_x$ does not lead to the changing of their sign. Only if a working region is symmetric with respect to the system center (in the *X*-direction), then the corresponding response remains to be an odd function of $B_x$ (Fig. 8). At the same time, comparing Fig. 7a and Fig. 7b, one can see that the symmetry relations associated with the inversion of $B_x$ are always satisfied only for the system as a whole. This fact clearly shows that we are truly dealing with an indivisible macroscopic quantum system where the behavior of its local regions cannot be treated separately from the behavior of the entire system.

### 3.2 Direct evidence for the quantum phase with macroscopic quantum orbits

Although the breaking of translational symmetry is a strong argument in favor of the quantum phase with macroscopic quantum orbits, to demonstrate this phase unambiguously we perform an ultimate experimental test. To clarify the core of the test, consider a thought experiment in Fig. 9. Suppose there is a rectangular quasi-one-dimensional macro-orbit in the system of disordered point-like charged defects. Let there are two distant local regions (shown in blue) where the scattering of a non-equilibrium electron can be detected. Suppose we inject an electron into the orbit by a strictly local photo-excitation of the region *A*. After that, the electron will undergo a local quasi-elastic scattering that result in a local current.

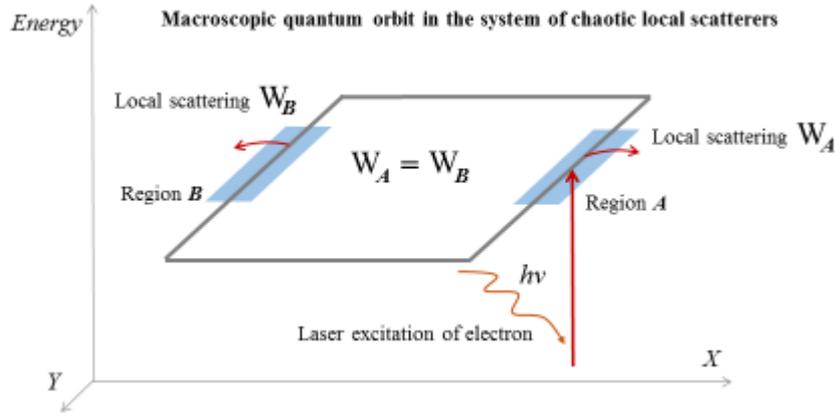

Fig. 9. Thought experiment to clarify the idea of an ultimate test for the presence of macro-orbits in the entire system.

But a "paradoxical" point is that, in accordance with the foundations of standard QM, electron fundamentally has no any definable position in a quantum orbit regardless of its lengthscale. This is due to the fact that the formalism of standard QM appeals to the so-called Hilbert space, i.e. to the space of quantum states, where any quantum orbit is a point that does not allow further fragmentation. As a result, the electron may undergo local scattering equiprobably at any point of the orbit. This means that, although electron was excited in region $A$, the probability to be scattered in this region ($W_A$) is exactly the same as the probability to be scattered in the region $B$ ($W_B$). Therefore, if the formalism of standard QM remains relevant in the macro-world, then the local photo-excitation of the region $A$ should lead to the same responses in both the region $A$ and the region $B$ regardless of the distance between them.

  Now if we come back to our experimental scheme (Fig. 5) and suppose that the spatial distribution of macro-orbits is such as in Fig. 2, then there are two families of working regions crossed by the same macro-orbits. The first one unifies four regions (No. 1, No. 2, No. 3, and No. 4) and the second one – two regions (No. 5 and No. 6). Thus, if, for example, we excite the system through window No. 5, then we should observe a photo-response not only in this region but also in the region No. 6 and moreover these responses should be almost the same despite the fact that the distance between them is five orders longer than the mean free path of excited electrons.

  But an even more non-trivial picture should be observed if we excite one of four working regions of the first family. In this case, the number of photo-electrons should be the same in all regions from No. 1 to No. 4 and hence photo-responses in these regions should also be the same, at least if we normalize them to the sensitivity of the system per one photo-electron, which is a function of the $X$-coordinate.

  Thus, we start with the experiment when all six windows are open and we measure responses in magnetic field of about 4.8 T in both the regime $A$ ($\alpha = 90°$) and the regime $B$ ($\alpha = 8°$) In this way, we find the sensitivity of each region in both regimes. As expected, in the regime $A$, the sensitivity is approximately the same in all regions whereas, in the regime $B$, the sensitivity is a function of the $X$-coordinate. After that, we close the windows one by one starting with the window No. 6 and ending with the window No. 1 and each time we measure responses in all working regions and normalize them to the local sensitivity.

  The outcome of the test is presented in the form of spatial diagrams of normalized responses for each configuration of local photo-excitation (Fig. 10). Here the height of parallelepipeds reflects the absolute value of responses while their position – the working region. If a response is shown in yellow, then the corresponding window is open. If it is shown in blue, then the window is closed.

**Closed windows: No. 6**

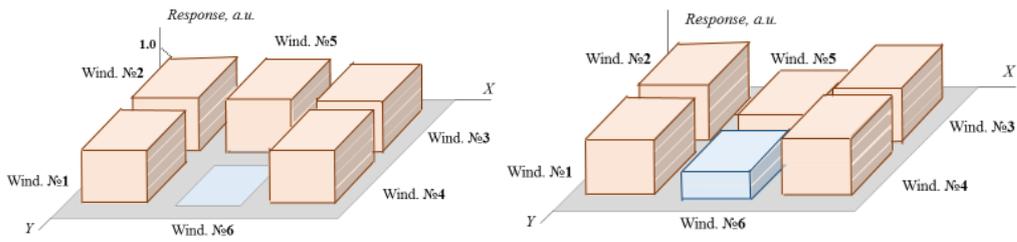

**Closed windows: No. 6, 5**

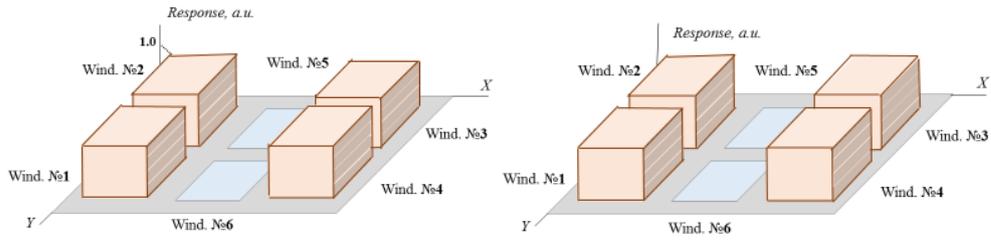

**Closed windows: No. 6, 5, 4**

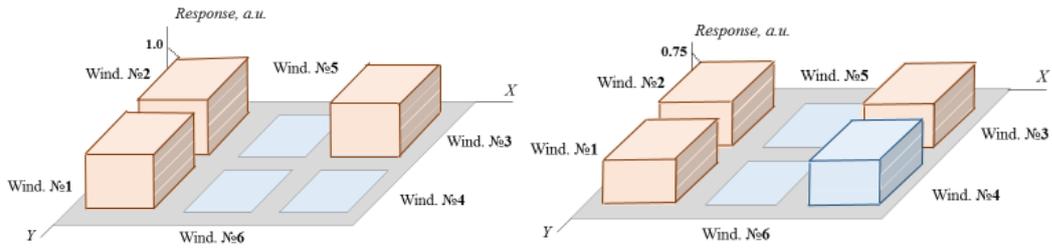

**Closed windows: No. 6, 5, 4, 3**

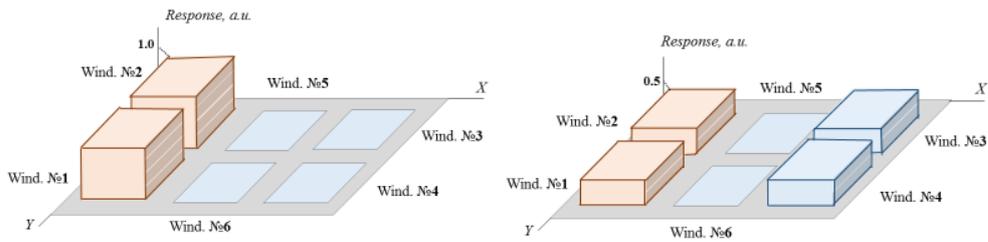

**Closed windows: No. 6, 5, 4, 3, 2**

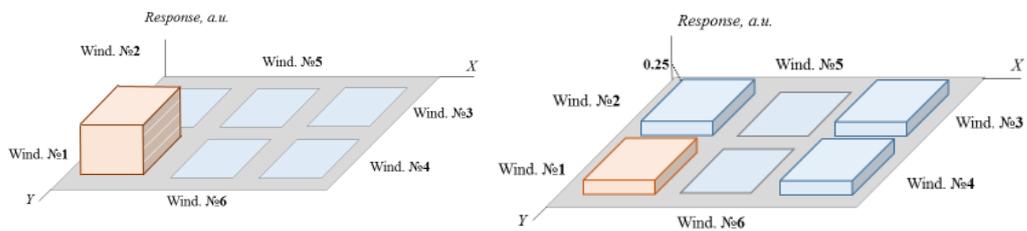

Fig. 10. Spatial diagrams of normalized responses under the different configurations of local photo-excitation. Magnetic field is about 4.8T. Left column – regime $A$ ($\alpha = 90°$); right column – regime $B$ ($\alpha = 8°$).

It is clearly seen that the behavior of responses does differ drastically in the regime A and in the regime B. In the former case, the behavior is quite trivial. Responses are observed only in the illuminated regions and their absolute value is always the same. This means that here the behavior of working regions is uncorrelated. By contrast, in the latter case, the behavior of responses seems astonishing at least in terms of our everyday intuition. For example, closing the window No. 6 does not eliminate the response in this region. Instead, we observe the same responses in the regions No. 5 and No. 6 as if they both are equally illuminated but with a halved intensity. But even more striking example of quantum correlations is the behavior of another family (from the region No. 1 to the region No. 4). It is clearly seen that here responses are always the same *regardless* of which of these regions are illuminated.

Thus, the photo-electrons do equiprobably emerge in any working region of the relevant family precisely in accordance with the prediction of standard QM and it seems that this fact leaves no room for any interpretation other than that in terms of the Halperin-type macro-orbits in the system interior. Nevertheless, to be sure that the emergence of photo-electrons far from the laser spot are not associated with any diffusion within the system, we perform a synchronous detection of responses in both illuminated and lightless regions from the same family. Fig. 11 shows typical tracks in both the region No. 1 and No. 3 ($\alpha = -8°$, $B = 4.8T$) when only the window No. 1 is open. As expected, the kinetics of both responses is truly similar to that of the laser pulse and there is no a delay between them at least within the time resolution.

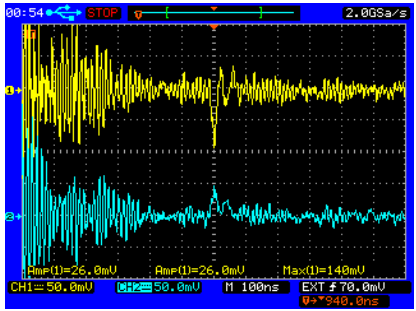

Fig. 11. Typical tracks under synchronous detection of responses in the region No. 1 (upper track) as well as in the region No. 3 (lower track) when only the window No. 1 is open ($\alpha = -8°$, $B = 4.8T$). Timescale is 100ns/div.

Thus, our experiments clearly indicate that if IQH system has an asymmetric confining potential, then a nonzero in-plane component of external magnetic field does lead to a quantum phase transition accompanied by the emergence of Halperin-type quasi-one-dimensional macro-orbits in the system interior. In this case, we get a macro-system, the local regions of which are in a strong quantum correlation so that such a system may be a unique tool to study the relevance of quantum laws at a macroscopic level. Various aspects of this study will be discussed elsewhere. Here we note only one of them related to such fundamental problem as the problem of interpretation of quantum theory.

The point is that, along with the standard QM, there are a number of alternative versions of quantum theory. One of them is the so-called de Broglie-Bohm pilot-wave theory [20]. The most attractive feature of this version is that, in contrast to standard QM, it allows one to interpret realistically almost all of the known quantum effects (for a review, see [21-22]). As a result, the de Broglie-Bohm theory had and still has quite a few supporters, among which, for example, is such outstanding physicist as John Bell [23].

As known, the main distinct feature of this theory is that here electrons always have both well-defined coordinates and well-defined velocity, so that their spatial dynamics can always be described in terms of trajectories known as Bohmian trajectories which differ drastically from the classical ones. On the contrary, standard QM does not support a fundamental description of electron motion in terms of trajectories [24]. However, despite such a principle difference, both versions successfully describe such non-trivial things as the double-slit experiment as well as the EPR correlations [25-26]. In fact, no experimental situations are currently known which would allow one to make an unambiguous choice in favor of one of these versions.

However, it may well be that our last experiment is precisely the situation of that kind. We mean the experiment where a significant part of photo-electrons emerges in the region No. 3 even though we excite only the region No. 1. This means these electrons somehow have covered a distance of about 1 cm. It is clear that such an electron transport cannot be interpreted in terms of any continuous trajectories (Bohmian or any others) because the distance covered is five orders longer than their mean free path. At the same time, this effect may well be described in terms of standard QM where electrons fundamentally have no any definable position in a quantum orbit regardless of its lengthscale and therefore their macroscopic transport through an intermediate orbit-like state should not necessarily be described in terms of a continuous spatial dynamics.

4. **Conclusions**

By the method of time-resolved terahertz laser spectroscopy, we provide strong evidence that in IQH systems a combined effect of an asymmetry of confining potential and a nonzero in-plane component of the external quantizing magnetic field may lead to a quantum phase transition with the emergence of Halperin-type quasi-one-dimensional macro-orbits in the system interior. We show experimentally that this transition is accompanied by the breaking of translational symmetry in the direction of the in-plane magnetic field whereas the macro-orbits are extended in the perpendicular direction. We also show that the local regions of the system are in a strong quantum correlation caused by the macro-orbits and the system is thus a unique tool to test the relevance of quantum laws in the macro-world. In particular, using such a system, we show experimentally that the description of macroscopic quantum spatial dynamics in terms of so-called Bohmian trajectories is untenable at least in a general case. At the same time, such dynamics may well be described in terms of standard QM where the concept of trajectory is not fundamental.

**Competing interests**

The author declares no competing interests.

**Acknowledgements**

The MBE samples were kindly provided by Prof. Sergey Ivanov (Ioffe Institute). The author thanks Prof. Raymond Chiao (UC at Merced) for useful comments on the experiment.